\documentclass[12pt,a4paper,oneside,final,tilepage,onecolumn,thmsal]{article}
\usepackage[dvips]{graphics}
\usepackage{epsfig}
\usepackage{amsmath}

\newcommand{\QED}{\hspace*{\fill}\rule{2.5mm}{2.5mm}}

\usepackage{graphics}
\begin{document}
\def\beq{\begin{equation}}
\def\eeq{\end{equation}}
\def\bea{\begin{eqnarray}}
\def\eea{\end{eqnarray}}
\def\ve{\vert}
\def\vel{\left|}
\def\ver{\right|}
\def\nnb{\nonumber}
\def\ga{\left(}
\def\dr{\right)}
\def\aga{\left\{}
\def\adr{\right\}}
\def\rar{\rightarrow}
\def\nnb{\nonumber}
\def\la{\langle}
\def\ra{\rangle}
\def\ba{\begin{array}}
\def\ea{\end{array}}
\def\tep{$B \rar K \ell^+ \ell^-$}
\def\tepm{$B \rar K \mu^+ \mu^-$}
\def\tept{$B \rar K \tau^+ \tau^-$}
\def\ds{\displaystyle}
\title{{\small {\bf Rare $B\rightarrow K^{*}\nu\bar{\nu}$ decay with polarized $K^{*}$ in
the fourth generation model }}}
\author{\vspace{1cm}\\
{\small T. BARAKAT} \thanks {electronic address:
barakat@ciu.edu.tr}\\ {\small Engineering Faculty, Cyprus
International University}\\ {\small Lefko\c{s}a, Mersin 10 -
Turkey } }
\date{}
\begin{titlepage}
\maketitle
\thispagestyle{empty}
\begin{abstract}
\baselineskip .8 cm The rare $B\rightarrow K^{*}\nu\bar{\nu}$
decay when $K^{*}$ meson is longitudinally or transversely
polarized is analysed in the context of the fourth generation
model. A significant enhancement to the missing energy spectrum
over the SM is recorded.
\end{abstract}
\vspace{1cm}
\end{titlepage}
\section{ Introduction}
\baselineskip .8cm \hspace{0.6cm} The theoretical and experimental
investigations of the rare decays has been a subject of continuous
interest in the existing literature. The experimental observation
of the inclusive $b\rightarrow X_{s}\gamma$ [1], and exclusive
$B\rightarrow K^{*}\gamma$ [2] decays, together with the recent
CLEO [3] upper limits on the exclusive decays $B \rightarrow
K^{*}\ell^{+}\ell^{-}$ which are less than one order of magnitude
above the SM predictions, stimulated the study of rare B meson
decays on a new footing. These decays take place via
flavor-changing neutral currents (FCNC) which are absent in the
Standard Model (SM) at tree level and appear only at the loop
level. The inclusive $B \rightarrow X_{s}\nu \bar{\nu}$ decay rate
is very sensitive to extensions of the SM, and provides a unique
source of constrains on some 'new physics' scenarios which predict
a large enhancement of this decay mode. Therefore, the study of $b
\rightarrow s\nu \bar{\nu}$, together with the search for $b
\rightarrow s\ell^{+} \ell^{-}$, and $ b\rightarrow s$ gluon
processes, with a refinement of the measurement of $B \rightarrow
X_{s}\gamma$ will allow to exploit a complete program to test the
SM properties at the loop level and constrain various new physics
scenarios.
 The first attempt to experimentally access the decay
 $b \rightarrow s\nu \bar{\nu}$ will be
through the exclusive modes, which will be better investigated at
B-factories. Among such modes, the channel $B \rightarrow
K^{*}\nu\bar{\nu}$ provokes special interest. The experimental
search for $B \rightarrow K^{*}\nu\bar{\nu}$ decays can be
performed through the large missing energy associated with the two
neutrinos, together with an opposite side fully reconstructed B
meson. The SM has been exploited to establish a bound on the
branching ratio of the above-mentioned decay of the order $\sim
10^{-5}$, which can be quite measurable for the upcoming $KEK$ and
SLAC B-factories. However, in SM there are three generations, and
yet, there is no theoretical argument to explain why there are
three and only three generations in SM, and there is neither an
experimental evidence for a fourth generation nor does any
experiment exclude such extra generations.
 On this basis, serious attempts to study the effects of the fourth generation
 on the rare B meson were made by many authors. For examples, the effects of
the fourth generation on the branching ratio of the $B \rightarrow
X_{s}\ell^{+}\ell^{-}$, and the $B \rightarrow X_{s}\gamma$ decays
is analysed in [4]. In [5] the fourth generation effects on the
rare exclusive $B \rightarrow K^{*}\ell^{+}\ell^{-}$ decay are
studied. In [6] the contributions of the fourth generation to the
$B_{s}\rightarrow \nu \bar{\nu}\gamma$ decay is investigated.
Recently, in [7] the effects of the fourth generation on the rare
$B\rightarrow K^{*}\nu\bar{\nu}$ decay is discussed.

In this work, the missing energy spectrum, and the branching ratio
of $B \rightarrow K^{*}\nu\bar{\nu}$ will be investigated when
$K^{*}$ meson is longitudinally or transversely polarized in a
sequential fourth generation model SM, which we shall call (SM4)
hereafter for the sake of simplicity. This model is considered as
natural extension of the SM, where the fourth generation model is
introduced in the same way the three generations are introduced in
the SM, so no new operators appear, and clearly the full operator
set is exactly the same as in SM. Hence, the fourth generation
will change only the values of the Wilson coefficients via virtual
exchange of a up-like quark $\acute{t}$. Subsequently, the missing
energy spectrum, and branching ratio of $B \rightarrow
K^{*}\nu\bar{\nu}$ are enhanced significantly, as we shall see, a
result which is in the right direction at least to help
experimental search for $B \rightarrow K^{*}\nu\bar{\nu}$ through
$m_{\acute{t}}$, and vice versa.

 Consequently, this paper is organized
as follows: in Section 2, the relevant effective Hamiltonian for
the decay $B\rightarrow K^{*}\nu\bar{\nu}$ in a sequential fourth
generation model (SM4) is presented; and in section 3, the
dependence of the missing energy spectrum, and branching ratio of
$B \rightarrow K^{*}\nu\bar{\nu}$ on the fourth generation model
parameters for the decay of interest is studied, when $K^{*}$
meson is longitudinally or transversely polarized using the
results of the Light- Cone QCD sum rules for estimating form
factors; and finally a brief discussion of the results is given.
\section{Effective Hamiltonian}
\hspace{0.6cm} In the Standard Model (SM), the process
$B\rightarrow K^{*}\nu\bar{\nu}$ is described at quark level by
the $b\rightarrow s\nu\bar{\nu}$ transition, and receives
contributions from Z-penguin and box diagrams, where dominant
contributions come from intermediate top quarks. The effective
Hamiltonian responsible for $b\rightarrow s\nu\bar{\nu}$ decay is
described by only one Wilson coefficient, namely $C^{(SM)}_{11}$,
and its explicit form is [8]:
\begin{eqnarray}
H_{eff}&=&\frac{G_{F}
\alpha}{2\pi\sqrt{2}sin^{2}\theta_{w}}C^{(SM)}_{11}
V^{*}_{ts}V_{tb}\bar{s}\gamma_{\mu}(1-\gamma_{5})b\bar{\nu}\gamma_{\mu}
(1-\gamma_{5})\nu,
\end{eqnarray}
where $G_{F}$ is the Fermi coupling constant, $\alpha$ is the fine
structure constant (at the Z mass scale), and $V^{*}_{ts}V_{tb}$
are products of Cabibbo-Kabayashi-Maskawa matrix elements. In
Eq.(1), the Wilson coefficient $C^{(SM)}_{11}$ in the context of
the SM has the following form including $O(\alpha_{s})$
corrections [9]:
\begin{eqnarray}
C^{(SM)}_{11}=\left[X_{0}(x_{t})+\frac{\alpha_{s}}{4\pi}X_{1}(x_{t})\right],
\end{eqnarray}
with
\begin{eqnarray}
X_{0}(x_{t})= \frac{x_{t}}{8}\left[\frac{x_{t}
+2}{x_{t}-1}+\frac{3(x_{t}-2)} {(x_{t}-1)^{2}}ln(x_{t})\right],
\end{eqnarray}
where $x_{t}=\frac{m^{2}_{t}}{m^{2}_{W}}$, and
\begin{eqnarray}
X_{1}(x_{t})&=&\frac{4x_{t}^{3}-5x_{t}^{2}-23x_{t}}{3(x_{t}-1)^{2}}-
\frac{x_{t}^{4}+x_{t}^{3}-11x_{t}^{2}+x_{t}}{(x_{t}-1)^{3}}ln(x_{t})+
\frac{x_{t}^{4}-x_{t}^{3}-4x_{t}^{2}-8x_{t}}{2(x_{t}-1)^{3}}ln^{2}(x_{t})
\nonumber \\
&+&\frac{x_{t}^{3}-4x_{t}}{(x_{t}-1)^{2}}Li_{2}(1-x_{t})+8x_{t}\frac{\partial
X_{0}(x_{t})}{\partial x_{t}} ln(x_{\mu}).
\end{eqnarray}
Here $Li_{2}(1-x_{t})=\int_{1}^{x_{t}}\frac{lnt}{1-t}dt$ is a
specific function, and $x_{\mu}=\frac{\mu^{2}}{m_{w}^{2}}$ with
$\mu=O(m_{t})$.

At $\mu=m_{t}$, the QCD correction for $X_{1}(x_{t})$ term is very
small (around $\sim 3\%$). From the theoretical point of view, the
transition $b\rightarrow s\nu\bar{\nu}$ is a very clean process,
since it is practically free from the scale dependence, and free
from any long distance effects. In addition, the presence of a
single operator governing the inclusive $b \rightarrow s
\nu\bar{\nu}$ transition is an appealing property. As has been
mentioned in the introduction, no new operators appear, and
clearly the full operator set is exactly same as in SM, thus the
fourth generation fermion changes only the values of the Wilson
coefficients $C^{(SM)}_{11}$ via virtual exchange of the fourth
generation up quark $\acute{t}$, i.e:
\begin{eqnarray}
C_{11}^{SM4}(\mu)&=&C^{(SM)}_{11}(\mu)+\frac{V^{*}_{\acute{t}s}V_{\acute{t}b}}
{V^{*}_{tb}V_{ts}}C^{(new)}(\mu),
\end{eqnarray}
where $C^{(new)}(\mu)$ can be obtained from $C^{(SM)}_{11}(\mu)$
by substituting $m_{t}\rightarrow m_{\acute{t}}$, and the last
terms in these expressions describe the contributions of the
$\acute{t}$ quark to the Wilson coefficients. $V_{\acute{t}s}$,
and $V_{\acute{t}b}$ are the two corresponding elements of the
$4\times 4$ Cabibbo-Kobayashi-Maskawa (CKM) matrix. In deriving
Eqs.(5) we factored out the term $V^{*}_{ts}V_{tb}$ in the
effective Hamiltonian given in Eq.(1).

As a result, we obtain a modified effective Hamiltonian, which
represents $b \rightarrow s \nu\bar{\nu}$ decay in the presence of
the fourth generation fermion:
\begin{eqnarray}
H_{eff}=\frac{G_{F}\alpha}{2\pi\sqrt{2}sin^{2}\theta_{w}}V^{*}_{ts}V_{tb}
[C_{11}^{(SM4)}]
\bar{s}\gamma_{\mu}(1-\gamma_{5})b\bar{\nu}\gamma_{\mu}
(1-\gamma_{5})\nu.
\end{eqnarray}
However, in spite of such theoretical advantages, it would be a
very difficult task to detect the inclusive $b \rightarrow s
\nu\bar{\nu}$  decay experimentally, because the final state
contains two missing neutrinos and many hadrons. Therefore, only
the exclusive channels, namely $B \rightarrow K^{*}(\rho)
\nu\bar{\nu}$, are well suited to search for, and constrain for
possible "new physics" effects. In order to compute $B \rightarrow
K^{*} \nu\bar{\nu}$ decay, we need the matrix elements of the
effective Hamiltonian Eq.(6) between the final, and initial meson
states. This problem is related to the non-perturbative sector of
QCD, and can be solved only by using non-perturbative methods. The
matrix element $<K^{*} \mid H_{eff}\mid B>$ has been investigated
in a framework of different approaches, such as chiral
perturbation theory [10], three point QCD sum rules [11],
relativistic quark model by the light front formalism [12],
effective heavy quark theory [13], and light cone QCD sum rules
[14,15]. To begin with, let us denote by $P_{B}$, and $P_{K^{*}}$
the four-momentum of the initial and final mesons, and define
q=$P_{B}-P_{K^{*}}$ as the four-momentum of the $\nu\bar{\nu}$
pair, and $x\equiv E_{miss}/M_{B}$ the missing energy fraction,
which is related to the squared four-momentum transfer $q^{2}$ by:
$q^{2}=M^{2}_{B}[2x-1+r^{2}_{K^{*}}]$, where $r_{K^{*}}\equiv
M_{K^{*}}/M_{B}$ with $M_{B}$, and $M_{K^{*}}$ being the initial
and final meson masses. The hadronic matrix element for the $B
\rightarrow K^{*} \nu\bar{\nu}$ can be parameterized in terms of
five form factors:
\begin{eqnarray}
<K^{*}_{h} \mid \bar{s}\gamma_{\mu}(1-\gamma_{5})b\mid B> =
\frac{2V(q^{2})}{M_{B}+M_{K^{*}}}
\epsilon_{\mu\nu\alpha\beta}\epsilon^{*\nu}(h)
P_{B}^{\alpha}P^{\beta}_{K^{*}}\nonumber
\\
 -i \left[\epsilon_{\mu}^{*}(h)(M_{B}+M_{K^{*}})A_{1}(q^{2})
-[\epsilon^{*}(h).q](P_{B}+P_{K^{*}})_{\mu}\frac{A_{2}(q^{2})}{M_{B}+M_{K^{*}}}
\right. \nonumber \\
 - \left. q_{\mu}[\epsilon^{*}(h).q]\frac{2M_{K^{*}}}{q^{2}}
[A_{3}(q^{2})-A_{0}(q^{2})] \right],
\end{eqnarray}
where  $\epsilon(h)$ is the polarization 4-vector of $K^{*}$
meson. The form factor $A_{3}(q^{2})$ can be written as a linear
combination of the form factors $A_{1}$ and $A_{2}$:
\begin{eqnarray}
A_{3}(q^{2})=\frac{1}{2M_{K^{*}}}\left[(M_{B}+M_{K^{*}})A_{1}(q^{2})-
(M_{B}-M_{K^{*}})A_{2}(q^{2})\right],
\end{eqnarray}
with a condition $A_{3}(q^{2}=0)=A_{0}(q^{2}=0)$.

From these form factors it is easy to derive the missing energy
distribution corresponding to the helicity  $h=0,\pm 1$of the
$K^{*}$ meson:
\begin{eqnarray}
\frac{d\Gamma(B \rightarrow K^{*}_{h=0} \nu\bar{\nu})}{dx}=
\frac{G_{F}^{2}\alpha^{2}M^{5}_{B}\mid
V^{*}_{ts}V_{tb}\mid^{2}}{64\pi^{5}sin^{4}\theta_{w}} \mid
C^{SM4}_{11}
\mid^{2}\frac{\sqrt{(1-x)^{2}-r^{2}_{K^{*}}}}{r^{2}_{K^{*}}(1+r^{2}_{K^{*}})^{2}}
\cdot \nonumber\\ \mid
(1+r^{2}_{K^{*}})^{2}(1-x-r^{2}_{K^{*}})A_{1}(q^{2})-
 2[(1-x)^{2}-r^{2}_{K^{*}}]A_{2}(q^{2}) \mid^{2},
\end{eqnarray}

\begin{eqnarray}
\frac{d\Gamma(B \rightarrow K^{*}_{h=\pm 1} \nu\bar{\nu})}{dx}=
\frac{G_{F}^{2}\alpha^{2}M^{5}_{B}\mid
V^{*}_{ts}V_{tb}\mid^{2}}{64\pi^{5}sin^{4}\theta_{w}} \mid
C^{SM4}_{11} \mid^{2}\sqrt{(1-x)^{2}-r^{2}_{K^{*}}}
\cdot\nonumber\\
 \frac{2x-1+r^{2}_{K^{*}}}{(1+r^{2}_{K^{*}})^{2}}
 \mid 2 \sqrt{(1-x)^{2}-r^{2}_{K^{*}}}V(q^{2})\mp (1+ r^{2}_{K^{*}})^{2}A_{1}(q^{2})
 \mid^{2}.
 \end{eqnarray}

From Eqs.(9,10), we can see that the missing energy spectrum for
$B \rightarrow K^{*} \nu\bar{\nu}$ contains three form factors: V,
$A_{1}$, and $A_{2}$. In this work, in estimating the missing
energy spectrum, we have used the results of [16]:
\begin{eqnarray}
F(q^{2})=\frac{F(0)}{1-a_{F}(q^{2}/M^{2}_{B})+b_{F}(q^{2}/M^{2}_{B})^{2}},
\end{eqnarray}
and the relevant values of the form factors at $q^{2}=0$ are:
\begin{eqnarray}
 A_{1}^{B \rightarrow K^{*}}(q^{2}=0)=0.34\pm 0.05,{~~} with{~~} a_{F}=0.6,
{~~}and{~~} b_{F}=-0.023,
\end{eqnarray}
\begin{eqnarray}
A_{2}^{B \rightarrow K^{*}}(q^{2}=0)=0.28\pm 0.04,{~~} with{~~}
a_{F}=1.18, {~~} and{~~} b_{F}=0.281,
\end{eqnarray}
and
\begin{eqnarray}
V^{B \rightarrow K^{*}}(q^{2}=0)=0.46\pm 0.07,{~~} with{~~}
a_{F}=1.55,{~~} and{~~} b_{F}=0.575.
\end{eqnarray}
 Note that all errors, which come out, are due to the uncertainties of the
b-quark mass, the Borel parameter variation, wave functions, and
radiative corrections are quadrature added in. Finally, to obtain
quantitative results we need the value of the fourth generation
CKM matrix elements $ V^{*}_{\acute{t}s}V_{\acute{t}b}$. For this
aim following [17], we will use the experimental results of the
decay $BR(B \rightarrow X_{s}\gamma)$ together with $BR(B
\rightarrow X_{c}e\bar{\nu_{e}})$ to determine the fourth
generation CKM factor $V^{*}_{\acute{t}s}V_{\acute{t}b}$. However,
in order to reduce the uncertainties arising from b-quark mass, we
consider the following ratio:
\begin{eqnarray}
R_{quark}=\frac{BR(B \rightarrow X_{s}\gamma)}{BR(B \rightarrow
X_{c}e\bar{\nu_{e}})}.
\end{eqnarray}
In the leading logarithmic approximation this ratio can be
summarized in a compact form as follows [18]:
\begin{eqnarray}
R_{quark}=\frac{\mid V^{*}_{ts}V_{tb} \mid ^{2}}{\mid V_{cb} \mid
^{2}}\frac{6\alpha}{\pi f(z)}  \mid C^{SM4}_{7}(m_{b}) \mid ^{2},
\end{eqnarray}
where
\begin{eqnarray}
f(z)=1-8z+8z^{3}-z^{4}-12z^{2}ln(z) {~~~~~~~} with {~~~}
z=\frac{m^{2}_{c,pole}}{m^{2}_{b,pole}}
\end{eqnarray}
is the phase space factor in $BR(B \rightarrow
X_{c}e\bar{\nu_{e}})$, and $\alpha= e^{2}/4\pi$. In the case of
four generation there is an additional contribution to $B
\rightarrow X_{s}\gamma$ from the virtual exchange of the fourth
generation up quark $\acute{t}$. The Wilson coefficients of the
dipole operators are given by:
\begin{eqnarray}
C^{SM4}_{7,8}(m_{b})=C^{SM}_{7,8}(m_{b})+\frac{
V^{*}_{\acute{t}s}V_{\acute{t}b}}
{V^{*}_{ts}V_{tb}}C^{new}_{7,8}(m_{b}),
\end{eqnarray}
where $C^{new}_{7,8}(m_{b})$ present the contributions of
$\acute{t}$ to the Wilson coefficients, and
$V^{*}_{\acute{t}s}V_{\acute{t}b}$ are the fourth generation CKM
matrix factor which we need now. With these Wilson coefficients
and the experiment results of the decays $BR(B \rightarrow
X_{s}\gamma)=2.66 \times 10^{-4}$, together with the semileptonic
$BR(B \rightarrow X_{c}e\bar{\nu_{e}})$=$0.103\pm 0.01$ [19,20]
decay, one can obtain the results of the fourth generation CKM
factor $ V^{*}_{\acute{t}s}V_{\acute{t}b}$, wherein, there exist
two cases, a positive, and a negative one [17]:
\begin{eqnarray}
( V^{*}_{\acute{t}s}V_{\acute{t}b}) ^{\pm}=\biggl[\pm \sqrt{
\frac{R_{quark} \mid V_{cb}\mid ^{2}\pi f(z)}{6\alpha \mid
V^{*}_{ts}V_{tb}\mid ^{2}}}-C^{(SM)}_{7}(m_{b}) \biggr] \frac{
V^{*}_{ts}V_{tb}}{C^{(new)}_{7}(m_{b})}.
\end{eqnarray}
The values for $V^{*}_{\acute{t}s}V_{\acute{t}b}$ are listed in
Table 1 [7].

A few comments about the numerical values of
$(V^{*}_{\acute{t}s}V_{\acute{t}b})^{\pm}$ are in order. From
unitarity condition of the CKM matrix we have
\begin{eqnarray}
V^{*}_{us}V_{ub}+V^{*}_{cs}V_{cb}+V^{*}_{ts}V_{tb}+V^{*}_{\acute{t}s}V_{\acute{t}b}=0.
\end{eqnarray}
If the average values of the CKM matrix elements in the SM are
used [19], the sum of the first three terms in Eq.(20) is about
$7.6\times 10^{-2}$. Substituting the value of
$(V^{*}_{\acute{t}s}V_{\acute{t}b})^{(+)}$ from Table 1 [7], we
observe that the sum of the four terms on the left-hand side of
Eq.(20) is closer to zero compared to the SM case, since
$(V^{*}_{\acute{t}s}V_{\acute{t}b})^{(+)}$  is very close to the
sum of the first three terms, but with opposite sign. On the other
hand if we consider $(V^{*}_{\acute{t}s}V_{\acute{t}b})^{-}$,
whose value is about $ 10^{-3}$, which is one order of magnitude
smaller compared to the previous case, and the error in sum of the
first three terms in Eq.(20) is about $\pm 0.6\times 10^{-2}$.
Therefore, it is easy to see then that, the value of
$(V^{*}_{\acute{t}s}V_{\acute{t}b})^{-}$ is within this error
range. In summary both $(V^{*}_{\acute{t}s}V_{\acute{t}b})^{+}$,
and $(V^{*}_{\acute{t}s}V_{\acute{t}b})^{-}$ satisfy the unitarity
condition of CKM, moreover, $\mid
(V^{*}_{\acute{t}s}V_{\acute{t}b})\mid ^{-} \leq
 10^{-1}\times \mid (V^{*}_{\acute{t}s}V_{\acute{t}b})\mid ^{+}$.
 Therefore, from our numerical analysis one cannot escape the conclusion
 that, the $(V^{*}_{\acute{t}s}V_{\acute{t}b})^{-}$ contribution to the
physical quantities should be practically indistinguishable from
SM results, and our numerical analysis confirms this expectation.
We now go on to put the above points in perspective.
\section{Numerical Analysis}
In order to investigate the sensitivity of the missing-energy
spectra, and branching ratios of rare $B \rightarrow K_{L}^{*}
\nu\bar{\nu}$, and $B \rightarrow K_{T}^{*} \nu\bar{\nu}$ decay
(where $K_{L}^{*}$, and $K_{T}^{*}$ stand for longitudinally and
transversely polarized $K^{*}$-meson, respectively)in SM4, the
following values have been used as input parameters:\\
$G_{F}=1.17{~}.10^{-5}~ GeV^{-2}$, $\alpha =1/137$, $m_{b}= 5.0$
GeV, $M_{B}= 5.28$ GeV, $\mid V^{*}_{ts}V_{tb}\mid$=0.045,
$M_{K^{*}}=0.892$ GeV, and the lifetime is taken as
$\tau(B_{d})=1.56\times 10^{-12}$ s [20], also  we have run
calculations of Eqs.(9,10) adopting the two sets of
$(V^{*}_{\acute{t}s}V_{\acute{t}b})^{\pm}$ in Table 1 [7]. we
present our numerical results for the missing-energy spectra, and
branching ratios in series of graphs. In figures (1-4), we show
the missing energy distribution to the decay $dBR(B \rightarrow
K_{L}^{*} \nu\bar{\nu})/dx$, and $dBR(B \rightarrow K_{T}^{*}
\nu\bar{\nu})/dx$ as functions of $x$; $
\frac{1-r^{2}_{K^{*}}}{2}\leq x \leq 1-r_{K^{*}}$, for
$m_{\acute{t}}$= 250 GeV, and $m_{\acute{t}}$= 350 GeV. It can be
seen their that, when $V^{*}_{\acute{t}s}V_{\acute{t}b}$ takes
positive values, i.e. $(V^{*}_{\acute{t}s}V_{\acute{t}b})^{-}$,
the missing energy spectrum is almost overlap with that of SM.
That is, the results in SM4 are the same as that in SM. But in the
second case, when the values of $V^{*}_{\acute{t}s}V_{\acute{t}b}$
are negative, i.e $(V^{*}_{\acute{t}s}V_{\acute{t}b})^{+}$ the
curve of the missing energy spectrum is quit different from that
of the SM. This can be clearly seen from figures (1-4). The
enhancement of the missing energy spectrum increases rapidly, and
the missing energy spectrum of the $K^{*}$ meson is almost
symmetrical. In figures (5,6), the branching ratio $BR(B
\rightarrow K_{L}^{*} \nu\bar{\nu})$, and $BR(B \rightarrow
K_{T}^{*} \nu\bar{\nu})$ are depicted as a function of
$m_{\acute{t}}$. Figures (5,6) show that for all values of
$m_{\acute{t}}\geq 210$ GeV the values of the branching ratios
become greater than SM. The enhancement of the branching ratio
increases rapidly with the increasing of $m_{\acute{t}}$. In this
case, the fourth generation effects are shown clearly. The reason
is that $(V^{*}_{\acute{t}s}V_{\acute{t}b})^{+}$ is 2-3 times
larger than $V^{*}_{ts}V_{tb}$ so that the last term in Eq.(5)
becomes important, and it depends on the $\acute{t}$ mass
strongly. Thus the effect of the fourth generation is significant.
Whereas, in our approach the predictions for the ratio $B
\rightarrow K_{L}^{*} \nu\bar{\nu}/B \rightarrow K_{T}^{*}
\nu\bar{\nu}$, as well as the transverse asymmetry $A_{T}$,
\begin{eqnarray}
A_{T}\equiv \frac{Br(B \rightarrow K_{h=-1}^{*} \nu\bar{\nu})-Br(B
\rightarrow K_{h=+1}^{*} \nu\bar{\nu})}{Br(B \rightarrow
K_{h=-1}^{*} \nu\bar{\nu})+Br(B \rightarrow K_{h=+1}^{*}
\nu\bar{\nu})}
\end{eqnarray}
are model-independent.

In conclusion, the missing-energy spectra, and branching ratio of
rare exclusive semileptonic $B \rightarrow K^{*} \nu\bar{\nu}$
decay has been investigated in the fourth generation model. The
effects of possible fourth generation fermion $\acute{t}$ quark
mass has been considered, and the sensitivity of the branching
ratio, and the missing-energy spectra to $\acute{t}$ quark mass is
observed.

 Finally, note that the results for $B\rightarrow
\rho\nu\bar{\nu}$ decay can be easily obtained from $B\rightarrow
K^{*}\nu\bar{\nu}$ when the following replacements are done in all
equations: $V_{tb}V^{*}_{ts}\rightarrow V_{tb}V^{*}_{td}$ and
$m_{K^{*}}\rightarrow m_{\rho}$. In obtaining these results, one
must keep in mind that the values of the form factors for
$B\rightarrow \rho$ transition generally differ from that of the
$B\rightarrow K^{*}$ transition. However, these differences must
be in the range of $SU(3)$ violation, namely in the order
$(15-20)\%$.

 \pagebreak

\pagebreak

 \begin{figure}
\centering
 \includegraphics[width=0.7\textwidth]{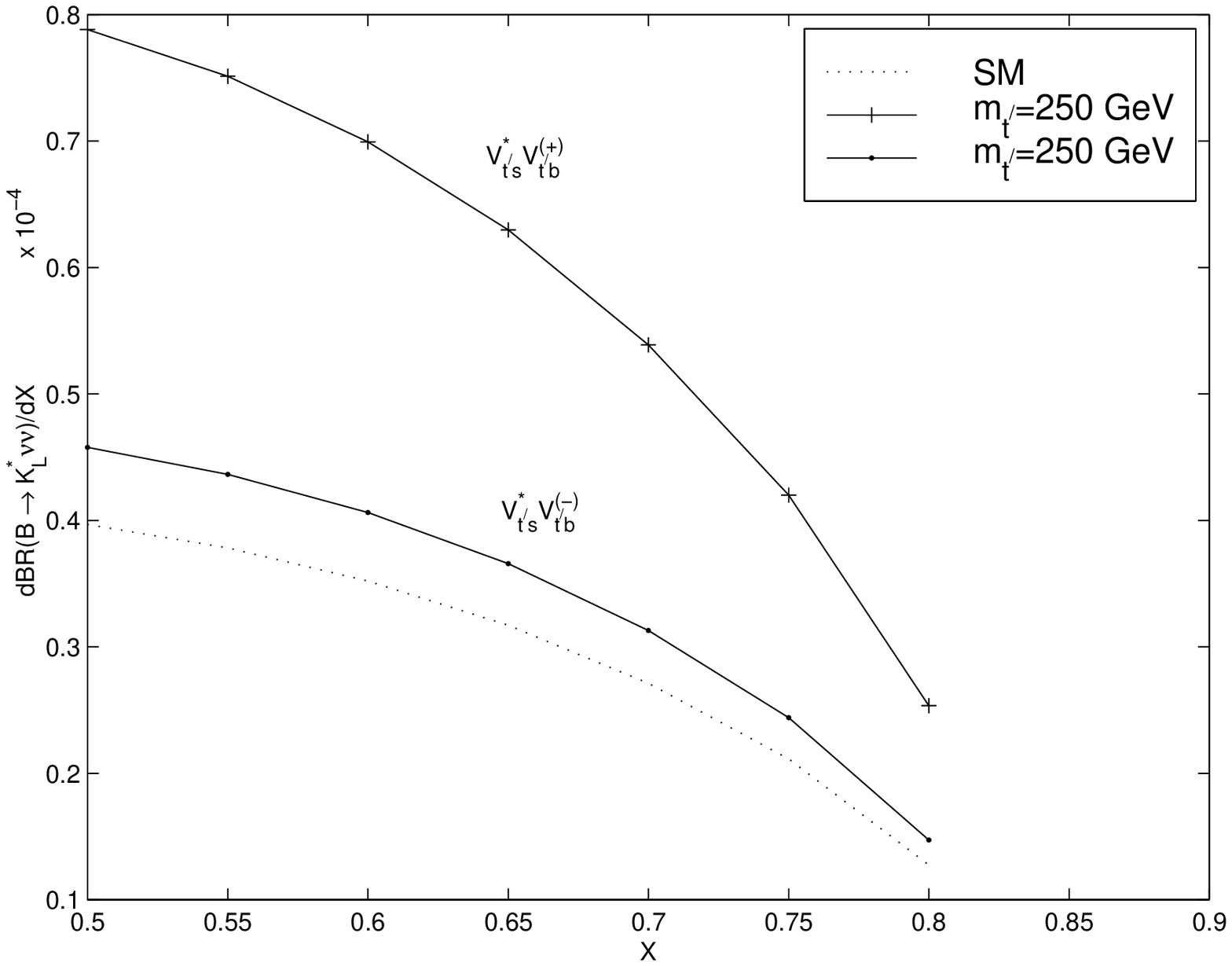}
 \caption{}
\label{fig 1}
 \end{figure}
\begin{figure}
\centering
 \includegraphics[width=0.7\textwidth]{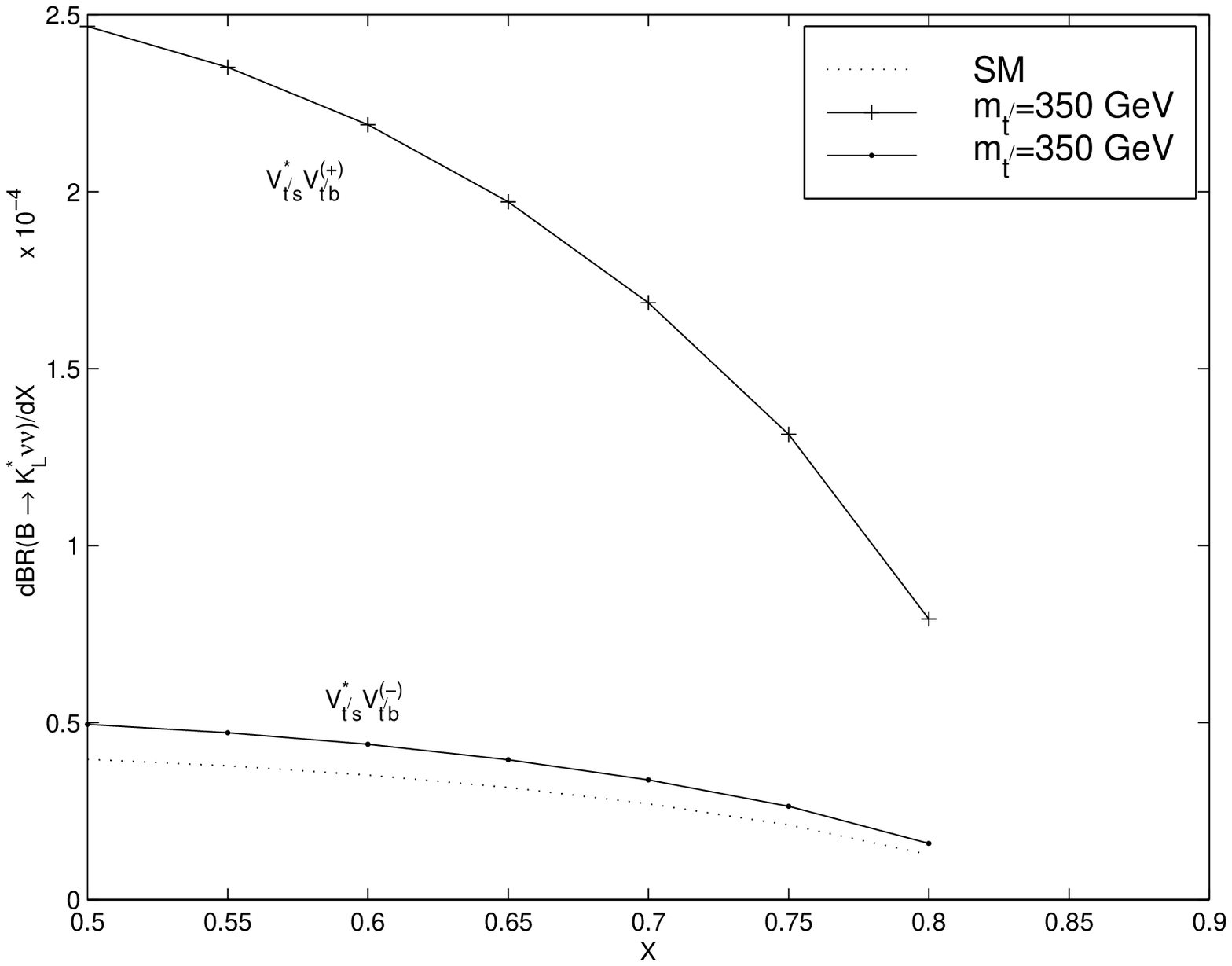}
 \caption{}
 \label{fig 2}
 \end{figure}
\begin{figure}
\centering
 \includegraphics[width=0.7\textwidth]{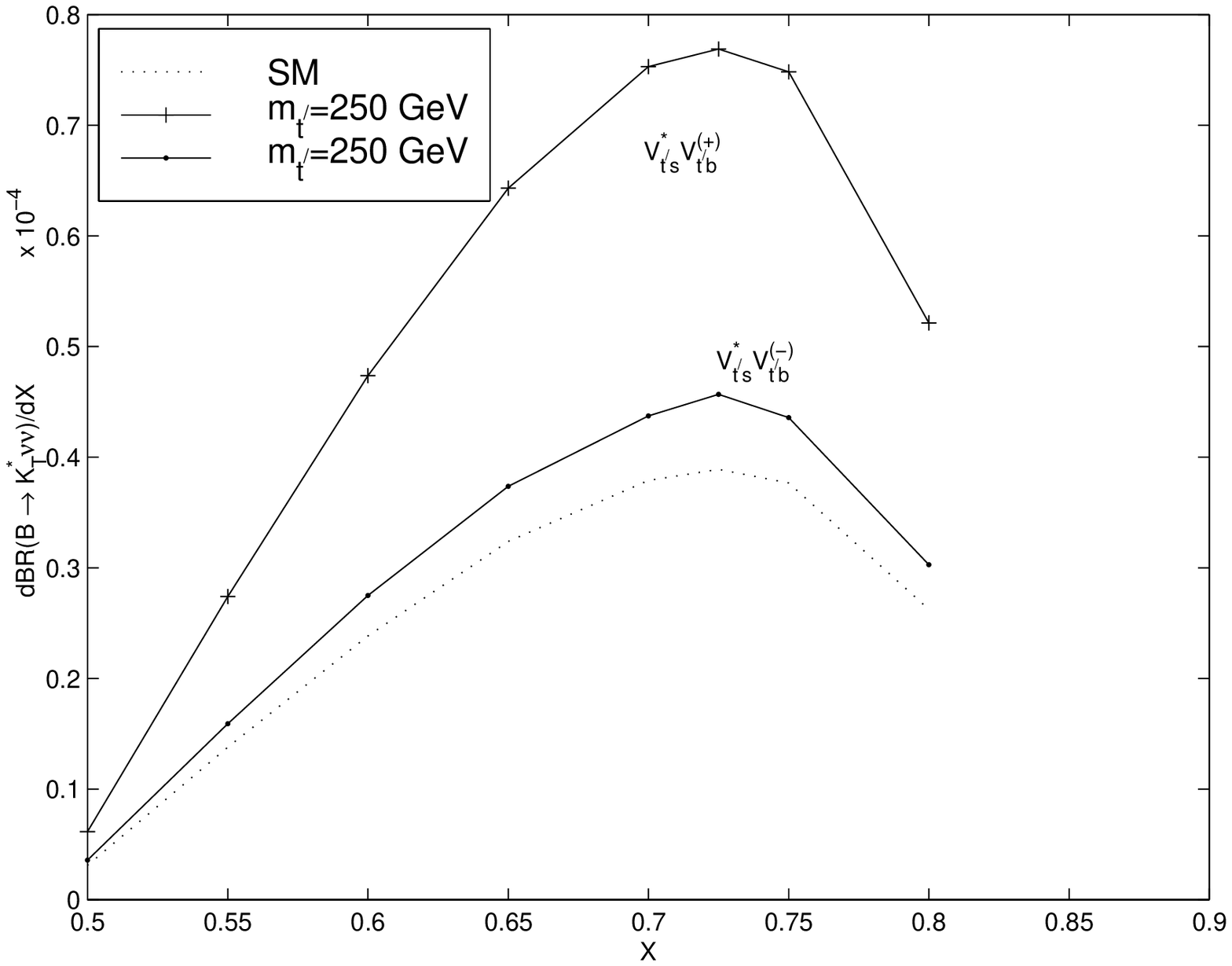}
 \caption{}
\label{fig 3}
 \end{figure}
 \begin{figure}
 \centering
 \includegraphics[width=0.7\textwidth]{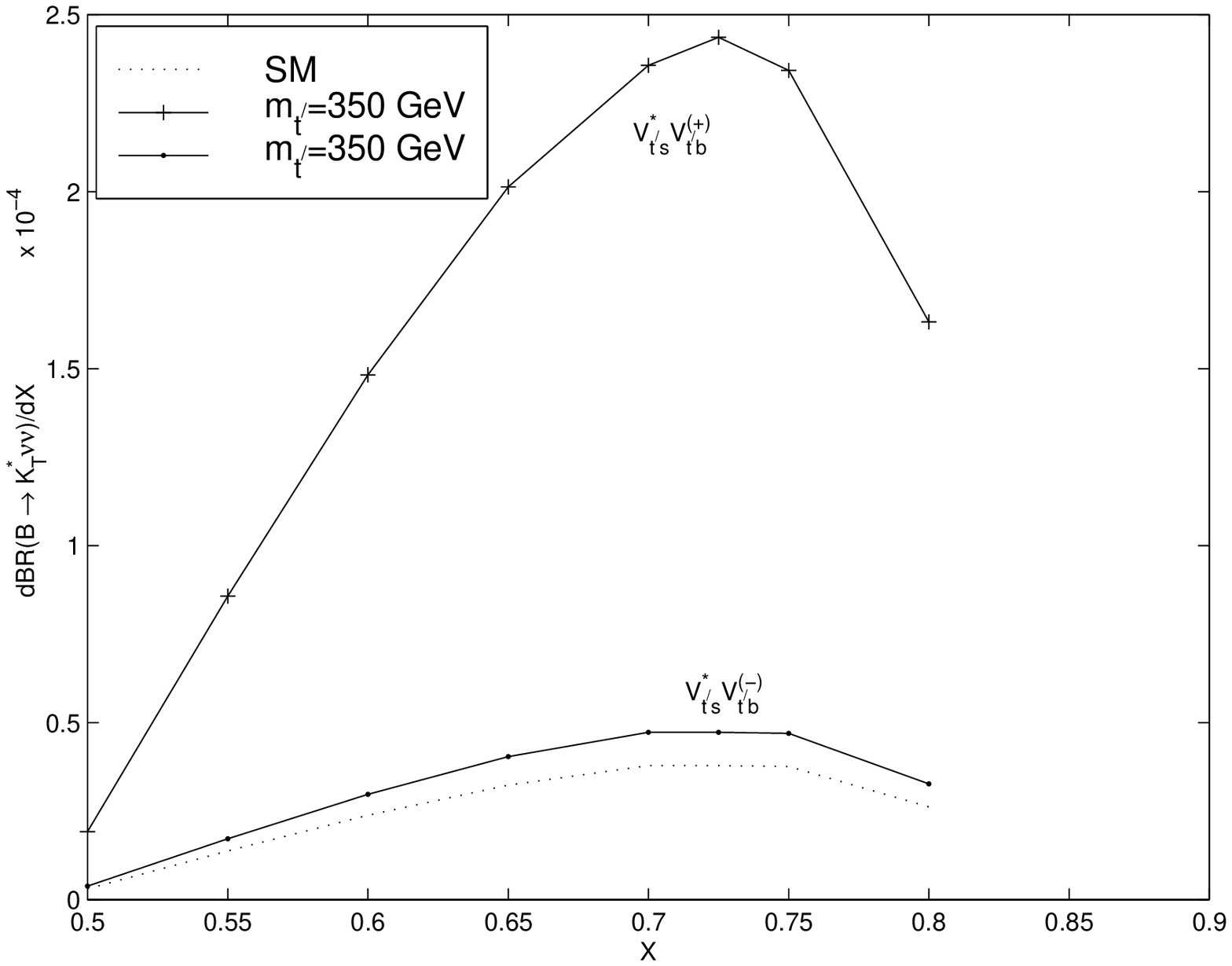}
 \caption{}
 \label{fig 4}
 \end{figure}
  \begin{figure}
 \centering
 \includegraphics[width=0.7\textwidth]{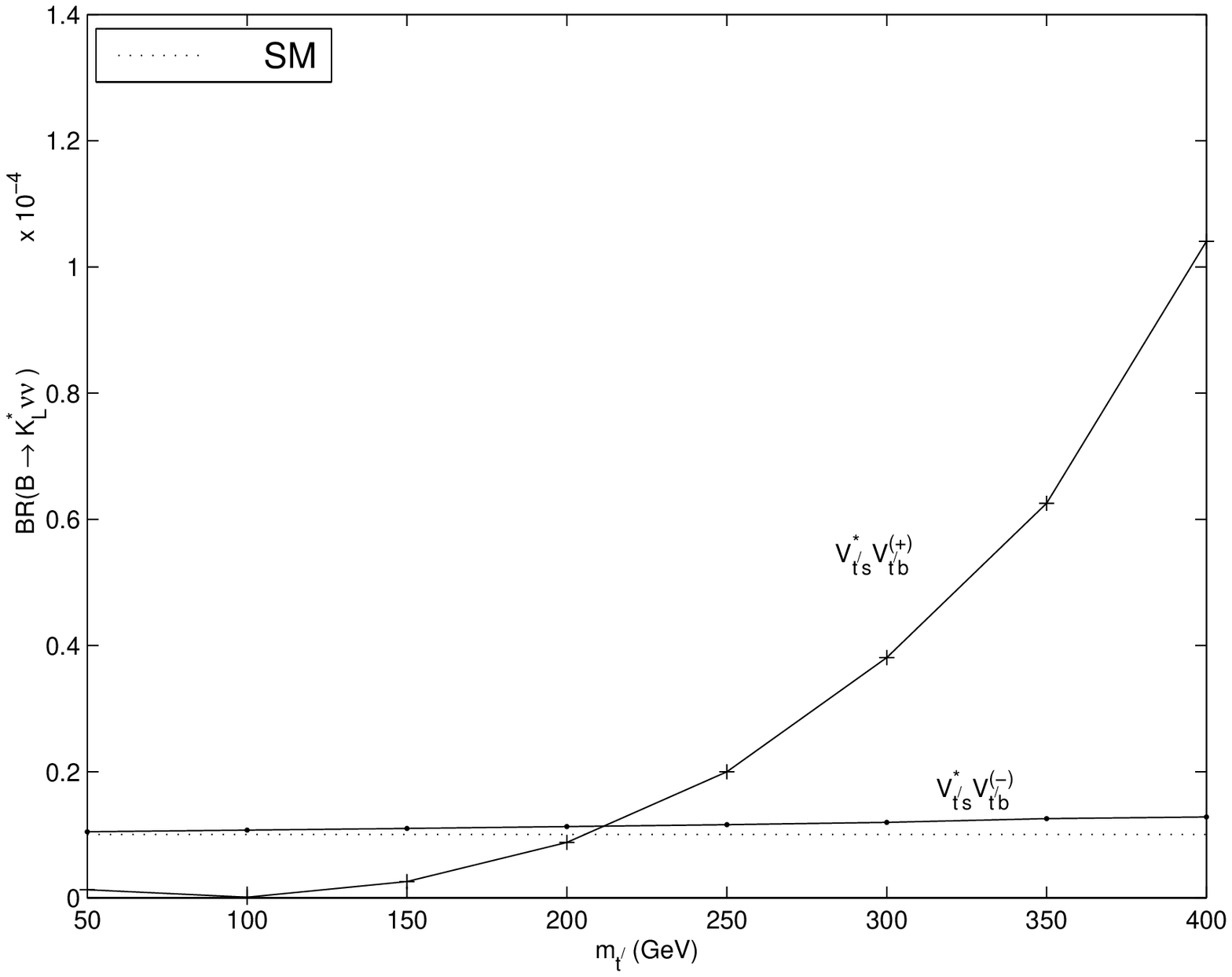}
 \caption{}
 \label{fig 5}
 \end{figure}
  \begin{figure}
 \centering
 \includegraphics[width=0.7\textwidth]{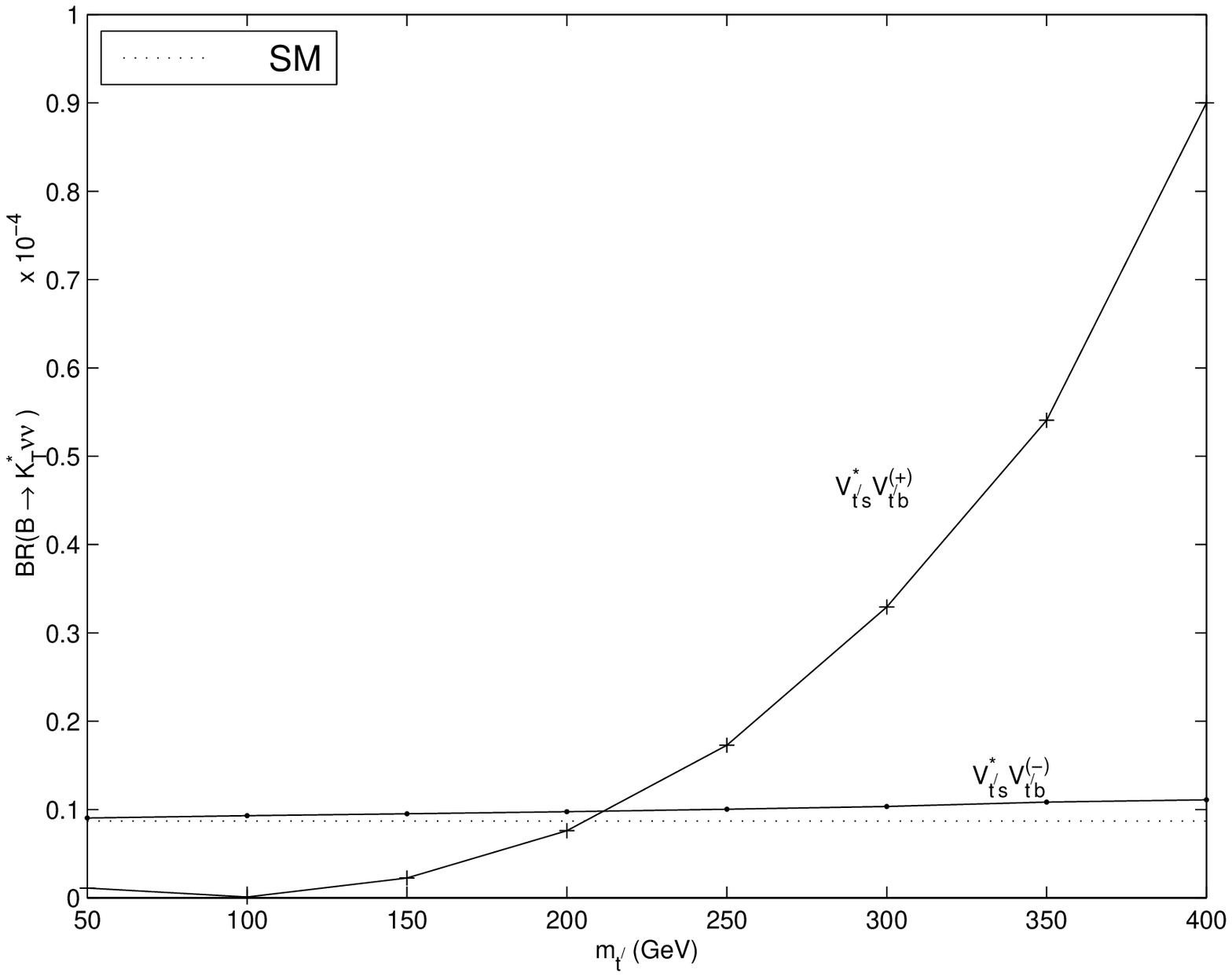}
 \caption{}
 \label{fig 6}
 \end{figure}

\begin{thebibliography}{99}
\bibitem{R1}M. S. Alam et al., CLEO Collaboration, Phys. Rev. Lett. 74
(1995) 2885.
\bibitem{R2}R. Ammar et al., CLEO Collaboration, Phys. Rev. Lett. 71
(1993) 674.
\bibitem{R3}S. Anderson et al., CLEO Collaboration, hep-ex/0106060 (2001).
\bibitem{R4}C.-S. Huang, W.-J. Huo, and Y.-L. Wu, Mod. Phys. A14
(1999) 2453, [hep-ph/9911203].
\bibitem{R5} T. M. Aliev, A. $\ddot{O}$zpineci, and M. Savci, Nucl. Phys. B (2000)
             275, [hep-ph/0002061].
\bibitem{R6} Y. Din\c{c}er, Phys. Lett. B505, (2001) 89, [hep-ph/0012135].
\bibitem{R7} T. Barakat, hep-ph/0105116 (2001).
\bibitem{R8} T. Barakat, J. Phys. G: Nucl.Part. Phys.24 (1998) 1903.
\bibitem{R9} G. Buchalla, and A. J. Buras, Nucl. Phys. B400 (1993) 225;
             G. Buchalla, A. J. Buras, and M. E. Lautenbacher,
             Rev. Mod. Phys. 68 (1996) 1125.
\bibitem{R10} R. Casalbuoni et al., Phys. Reports 281 (1997) 145.
\bibitem{R11} P. Colangelo, F. De Fazio, P. Santorelli, and E. Scrimieri, Phys.
             Rev. D53 (1996) 3672.
\bibitem{R12} W. Jaus, and D. Wyler, Phys. Rev. D41 (1991) 3405; D. Melikhov,
             N. Nikitin, and S. Simula, Phys. Lett. B410, (1997) 290, [hep-ph/9704268].
\bibitem{R13} W. Roberts, Phys. Rev. D54 (1996) 863.
\bibitem{R14} T. M. Aliev, A. $\ddot{O}$zpineci, and M. Savci, Phys.Rev. D5 (1996)
             4260.
\bibitem{R15} P. Ball, and V. M. Braun, Phys. Rev. D55 (1997) 5561.
\bibitem{R16} P. Ball, Fermilab-Conf-98/098-T, [hep-ph/9803501];
P. Ball, and V. M. Braun, Phys. Rev. D58 (1998) 094016.
\bibitem{R17} W.-J. Huo, [hep-ph/0006110].
\bibitem{R18} A. J. Buras, TUM-hep-316/98, [hep-ph/9806471].
\bibitem{R19}M. S. Alam, Phys. Rev. Lett. 74 (1995) 2885.
\bibitem{R20}C. Caso et al., Particle Data Group, Eur.Phys. J. C3 (1998) 1.
\end{thebibliography}
\end{document}